\documentclass[
superscriptaddress,
amsmath,amssymb,
aps,
prmaterials,
twocolumn
]{revtex4-2}
\usepackage{graphicx}
\usepackage{xcolor}

\begin{document}

\title{Quantifying the Kinetic and Thermodynamic Origins of \\
the Glass-Forming Ability in Metals}

\author{Weiwei Jin}
\thanks{These authors contributed equally to this work.}
\affiliation{Department of Mechanical Engineering, Yale University, New Haven, Connecticut, 06520, USA}
\author{Jinpeng Fan}
\thanks{These authors contributed equally to this work.}
\affiliation{Department of Mechanical Engineering, Yale University, New Haven, Connecticut, 06520, USA}
\author{Mark D. Shattuck}
\affiliation{Benjamin Levich Institute and Physics Department, The City College of The City University of New York, New York, New York 10031, USA}
\author{Corey S. O'Hern}
\thanks{Contact author: corey.ohern@yale.edu}
\affiliation{Department of Mechanical Engineering, Yale University, New Haven, Connecticut, 06520, USA}
\affiliation{Department of Applied Physics, Yale University, New Haven, Connecticut, 06520, USA}
\affiliation{Department of Materials Science, Yale University, New Haven, Connecticut, 06520, USA}
\affiliation{Department of Physics, Yale University, New Haven, Connecticut, 06520, USA}

\begin{abstract}
The glass-forming ability (GFA) of a metallic liquid can be quantified by the critical cooling rate $R_c$, below which it crystallizes upon cooling. $R_c$ is determined by the competition between $R$ and the crystallization rate $\mathcal{K}_{\rm X}$. Here, we calculate $R_c$ and $\mathcal{K}_{\rm X}$ in model pure metals and alloys with BCC/B2 and FCC equilibrium crystal structures. Using classical nucleation theory and independently calculated crystal growth rates, we evaluate $\mathcal{K}_{\rm X}=(IU^3)^{1/4}$ at $T=0.8T_m$, where $I$ and $U$ are the nucleation and growth rates and $T_m$ is the melting temperature. We find a strong positive correlation between $\mathcal{K}_{\rm X}$ and $R_c$ over eleven orders of magnitude. We decompose $\mathcal{K}_{\rm X}$ into a product of a kinetic contribution, involving atomic cluster attachment and diffusion, and a thermodynamic contribution. We show that the thermodynamic contribution to $\mathcal{K}_{\rm X}$ dominates the kinetic contribution over the range of cooling rates that we consider. The nucleation barrier $\Delta G_c \sim \tilde\gamma^3/\Delta\tilde\mu^2$, where $\tilde\gamma$ is the reduced effective liquid--crystal interfacial free energy and $\Delta\tilde\mu$ is the reduced bulk driving force, determines the magnitude of the thermodynamic contribution. The relative importance of $\tilde\gamma$ and $\Delta\tilde\mu$ depends on the material: a large interfacial free energy can suppress nucleation in some systems, whereas a weak bulk driving force can inhibit crystallization in others. Analyses of the structural and chemical order show that better glass formers have enhanced structural frustration, more icosahedral-like and less crystal-like local environments, and a larger chemical-order mismatch between the supercooled liquid and crystal phases. In this work, we have distinguished the kinetic and thermodynamic contributions to the GFA in metallic liquids and identified multiple thermodynamic mechanisms that can improve glass formation.
\end{abstract}

\maketitle

\section{Introduction}
\label{sec:intro}

Metallic liquids form glasses with amorphous atomic structures when they are cooled rapidly below the melting point to bypass crystallization. They can be used in numerous industrial applications due to their high strength, large elastic limit, and corrosion resistance~\cite{wang2004bulk, schroers2013bulk,gao2022recent,kotni2026review,jin2023general}. The glass-forming ability (GFA) can be characterized via the critical cooling rate $R_c$, which is the minimum cooling rate above which crystallization does not occur. Materials with smaller $R_c$ indicate better glass-forming ability. $R_c$ can vary by more than $13$ orders of magnitude between pure metals and bulk metallic glasses. Understanding the key physical features that control the GFA would enable accurate predictions of the GFA for alloys~\cite{turnbull1969conditions,kelton2010nucleation,hu2020glass,hu2022glass,liu2024effective}.

Inoue's empirical rules for good GFA, including multicomponent compositions, $> 10\%$ atomic-size mismatch among elements, and negative mixing enthalpies, have guided the development of bulk metallic glasses~\cite{inoue2000stabilization,wang2004bulk}. 
Although these guidelines have enabled the discovery of numerous glass-forming alloys, they do not provide a quantitative framework for predicting the GFA. Glass formation requires crystal nucleation and subsequent growth to remain sufficiently slow over the temperature interval that is traversed during cooling~\cite{turnbull1969conditions,kelton2010nucleation}. 
Within classical nucleation theory (CNT), crystal nucleation depends on the atomic attachment rate and the free-energy barrier determined jointly by the liquid--crystal interfacial free energy and the chemical-potential difference (i.e., the bulk driving force) between the supercooled liquid and crystal phases~\cite{avramov2007kinetics,zaragoza2015competition,houghton2025sample}. Crystal growth likewise depends on atomic transport and the thermodynamic driving force for crystallization~\cite{kelton2010nucleation}. Developing the ability to predict the GFA of alloys therefore requires calculating the nucleation and growth rates and determining their kinetic and thermodynamic contributions.

Molecular dynamics (MD) simulations that cool metallic liquids below their melting temperatures can be used to directly calculate the critical cooling rate for $R_c > 10^7 {\rm K/s} $. A complementary approach for determining the GFA is to calculate the crystal nucleation $I$ and growth rates $U$ at a representative undercooled temperature to obtain the crystallization rate $\mathcal{K}_{\rm X}=(IU^3)^{1/4}$ given homogeneous steady-state nucleation and isotropic three-dimensional growth~\cite{avramov2007kinetics,houghton2025sample}. Glass formation will typically occur when $\frac{R}{T_m}/{\mathcal K}_{\rm X} \gg 1$, and thus the GFA can be estimated using $R_c/T_m \approx \mathcal{K}_{\rm X}$. 

Recent studies have shown that the liquid--crystal interfacial free energy can strongly influence the nucleation barrier and, consequently, the GFA of metallic liquids (such as Cu-Zr, Ni-Al, and Zr) that crystallize into body-centered cubic (BCC/B2) structures~\cite{hu2020physical}. 
Subsequent work has emphasized that crystal-like preordering in supercooled liquids can promote crystallization by decreasing the liquid--crystal interfacial free energy~\cite{hu2022revealing}. These findings suggest that local structural and chemical ordering can affect the formation of crystal nuclei. However, is the interfacial free energy the dominant contribution to the nucleation barrier across a broader range of alloys and crystal symmetries, or does the driving force also contribute significantly to the free energy barrier? In particular, a large nucleation barrier that inhibits crystallization may arise from either a high interfacial free energy cost or a weak bulk driving force. 

In this work, we address this important question using embedded atom method (EAM) models of $12$ pure metals and alloys with BCC/B2 and FCC equilibrium crystal structures. First, we calculate the nucleation rate from the attachment rate, critical nucleus size, and the driving force using CNT and obtain the growth rate directly from MD simulations of the liquid--solid interface at temperature $T=0.8T_m$. We find that $\mathcal{K}_{\rm X}$ and $R_c$ are strongly correlated over $\sim 11$ orders of magnitude. 
Second, we decompose $\mathcal{K}_{\rm X}$ into the product of a kinetic and a thermodynamic contribution to determine which factor controls the crystallization rate. We show that the thermodynamic contribution varies significantly more than the kinetic contribution over the range of $R_c$ that we studied. 
Third, we identify the distinct thermodynamic mechanisms for suppressing crystallization in each metallic liquid: a large effective liquid--crystal interfacial free energy prevents crystallization in some systems, whereas a weak bulk driving force inhibits crystallization in others. These findings highlight the importance of understanding both the interfacial and bulk thermodynamic contributions to the crystallization rate. Finally, we examine structural ordering in the supercooled liquid and the chemical-order mismatch between the liquid and crystal as possible microscopic mechanisms that explain the high liquid--crystal interfacial free energy and weak driving force that occur in glass-forming systems over a range of $R_c$. 

The remainder of the article is organized as follows. In Sec.~\ref{sec:method}, we describe the EAM models for the $12$ pure metals and alloys that we study and the MD simulations used to calculate $R_c$ for each system. In Sec.~\ref{sec:results}, we show the correlation between $R_c$ and the crystallization rate $\mathcal{K}_{\rm X}$ obtained by calculating the nucleation and growth rates. We also decompose $\mathcal{K}_{\rm X}$ into the kinetic and thermodynamic contributions. We show that the dominant thermodynamic contribution to $\mathcal{K}_{\rm X}$ arises from the nucleation barrier, which is determined by the liquid--crystal interfacial free energy and bulk driving force. We further use metrics of local structural and chemical order to provide microscopic mechanisms that give rise to changes in the nucleation barrier. 
In Sec.~\ref{sec:conclusions}, we summarize the main findings of this work and discuss promising future research directions, including the temperature dependence of crystallization kinetics, anisotropy of the liquid-crystal interface, and a wider range of alloy compositions and crystal symmetries. Four appendices are included to provide additional details of the computational methods. In Appendix~\ref{app:Tm}, we describe calculations of the melting temperature and enthalpy of fusion. In Appendix~\ref{app:order}, we describe the metrics used to characterize the local structural and chemical order. Appendix~\ref{app:seed} outlines the seeding method for evaluating the atomic attachment rate and critical nucleus. Finally, Appendix~\ref{app:growth} describes the MD simulations used to calculate the crystal growth rates. 

\section{Methods}
\label{sec:method}

We investigate $12$ pure metals and binary alloys with different equilibrium crystal structures using EAM interatomic potentials. The BCC systems include Ti~\cite{mendelev2016development}, Zr~\cite{mendelev2009development}, Fe~\cite{mendelev2003development}, and Nb~\cite{fellinger2010force}.
We also consider CuZr~\cite{mendelev2009development} and NiAl~\cite{mishin2002embedded}, which crystallize into the B2 structure, consisting of two interpenetrating simple-cubic sublattices that together form a BCC lattice. The FCC systems include pure Ni and Al~\cite{mishin2002embedded}, Cu~\cite{mendelev2009development}, and Ni$_9$Al and Ni$_{17}$Al$_3$ alloys~\cite{mishin2002embedded}. For the latter two alloys, the crystal structures are chemically disordered FCC lattices, with the constituent elements randomly distributed over the lattice sites. Finally, we include $\gamma$-TiAl~\cite{zope2003interatomic}, which forms a tetragonal L$1_0$ structure, a tetragonally distorted FCC lattice. 

To determine the critical cooling rate $R_c$, we perform MD simulations of systems containing $N=2000$ atoms, initially arranged on a BCC lattice with periodic boundary conditions in all three directions. This system size is selected based on previous assessments of weak finite-size effects in $R_c$ calculations when $N>10^3$~\cite{hu2019tuning,hu2020glass}. Each system is heated to a temperature of $1500$ K above its melting temperature $T_m$ (determined as described in Appendix~\ref{app:Tm}) and equilibrated in the liquid state. The liquid is then quenched to room temperature at a prescribed cooling rate $R$. 
The simulations are conducted in the isothermal--isobaric ($NpT$) ensemble at zero external pressure $p$, using a Nos\'e--Hoover thermostat and barostat. The equations of motion are integrated using a modified velocity-Verlet algorithm with time step $\Delta t = 0.002$ ps. 

After cooling to room temperature, we determine the fraction of atoms in a locally crystalline environment $F_{\rm xtal}$ using the structural criteria described in Appendix~\ref{app:order}. The dependence of $F_{\rm xtal}$ on $R$ is fitted to
\begin{equation}
    F_{\rm xtal}=\frac{1}{2}\left\{1-\tanh\left[\frac{1}{\kappa}\log_{10}\left(R/R_c \right)\right]\right\},
    \label{eq:Rc_fit}
\end{equation}
where $0<\kappa<1$ is the stretching exponent~\cite{hu2022glass}. By construction, $R_c$ is the cooling rate at which $F_{\rm xtal}=0.5$. For CuZr, we use the experimentally determined $R_c=250$ K/s~\cite{wang2005understanding,wang2004CuZr}, while for the other systems, $R_c$ is obtained from the MD simulations of cooling at different rates.

To characterize the crystallization kinetics, we determine the critical nucleus size $N_c$ and atomic attachment rate $f^+$ using the seeding method (see Appendix~\ref{app:seed}). The melting temperature $T_m$ and enthalpy of fusion $\Delta H_m$ used to calculate the thermodynamic driving force $\Delta\mu$ are obtained using standard methods described in Appendix~\ref{app:Tm}. Thus, the crystal nucleation rate can be obtained from classical nucleation theory using $N_c$, $f^+$, and $\Delta\mu$. We calculate the crystal growth rate directly from MD simulations using planar liquid--crystal interfaces as shown in Appendix~\ref{app:growth}.

\section{Results}
\label{sec:results}

The value of $R_c$ for metallic liquids reflects the competition between crystallization and vitrification during cooling. In this section, we first show the strong correlation between the crystallization rate $\mathcal{K}_{\rm X}$ and $R_c$ for $12$ metallic liquids with different equilibrium crystal structures. 
We quantify the kinetic and thermodynamic contributions to $\mathcal{K}_{\rm X}$, show that the thermodynamic contribution is dominant, and then identify which thermodynamic factors enhance the nucleation barrier in each system. Finally, we examine the structural and chemical ordering in the supercooled liquid and crystal phases to investigate the microscopic mechanisms that determine the nucleation barrier.

\subsection{Correlation between the crystallization rate and critical cooling rate}
\label{sec:resultA}

To relate $R_c$ to the microscopic crystallization kinetics, we calculate a characteristic crystallization rate that combines nucleation and growth processes. 
According to Johnson--Mehl--Avrami--Kolmogorov (JMAK) kinetics~\cite{avramov2007kinetics,houghton2025sample} for homogeneous, steady-state nucleation and isotropic three-dimensional growth, the time $t$ required to reach a crystallized volume fraction $\varphi$ during isothermal crystallization is 
\begin{equation}
    t=\left[\frac{-3\ln(1-\varphi)}{\pi I U^3}\right]^{1/4},
    \label{eq:JMAK}
\end{equation}
where $I$ is the crystal nucleation rate per unit volume and $U$ is the crystal growth rate. We therefore define the characteristic crystallization rate, 
\begin{equation}
    \mathcal{K}_{\rm X}=(IU^3)^{1/4},
    \label{eq:KX}
\end{equation}
which has units of inverse time.

Within classical nucleation theory, the steady-state crystal nucleation rate of a supercooled liquid is
\begin{equation}
    I(T)=\rho_l Z f^+ \exp\left(- \beta\Delta G_c\right),
    \label{eq:I}
\end{equation}
where $\rho_l$ is the liquid number density, $f^+$ is the attachment rate of atoms to the critical nucleus, $\beta = 1/(k_{\textrm B} T)$, and $k_{\textrm B}$ is the Boltzmann constant. The Zeldovich factor for a spherical crystal nucleus is 
\begin{equation}
    Z=\sqrt{\frac{\beta\Delta \mu}{6\pi N_c}},
    \label{eq:Z}
\end{equation}
where  $\Delta \mu=\mu_l-\mu_s>0$ is the chemical-potential difference between the liquid and crystal phases and $N_c$ is the number of atoms in the critical nucleus. $Z$ characterizes the curvature of the nucleation free-energy barrier at the critical nucleus size. As shown in Sec.~\ref{sec:resultC}, $Z$ does not vary significantly over the $12$ metallic liquids that we studied. The corresponding nucleation free-energy barrier is
\begin{equation}
    \Delta G_c=\frac{N_c\Delta \mu}{2}.
\end{equation}

The crystal growth rate is described using the classical Wilson–Frenkel relation~\cite{kelton2010nucleation,wilson1900on,frenkel1932note},
\begin{equation}
    U(T)=k_0 D\left[1-\exp\left(-\beta\Delta\mu\right)\right],
    \label{eq:U}
\end{equation}
where $k_0$ is a kinetic prefactor and $D$ is the diffusion coefficient of the slower alloy component.  
Thus, the nucleation and growth processes both depend on atomic transport and the thermodynamic driving force for crystallization.

Here, we evaluate the nucleation and growth rates at $T=0.8T_m$ to compare systems at the same reduced undercooling.
The critical nucleus size $N_c$ and attachment rate $f^+$ are obtained using the seeding method, while the crystal growth rate is calculated directly from MD simulations that employ a planar liquid--crystal interface. The chemical-potential difference is approximated by $\Delta\mu(T)=\Delta H_m\left(1-{T}/{T_m}\right)$,
where $\Delta H_m$ is the enthalpy of fusion per atom at $T_m$. This approximation neglects the temperature dependence of the enthalpy difference during the cooling process. Details of these calculations are provided in Appendices~\ref{app:Tm}, \ref{app:seed}, and \ref{app:growth}.

\begin{figure}[t]
    \centering
    \includegraphics[width=0.95\linewidth]{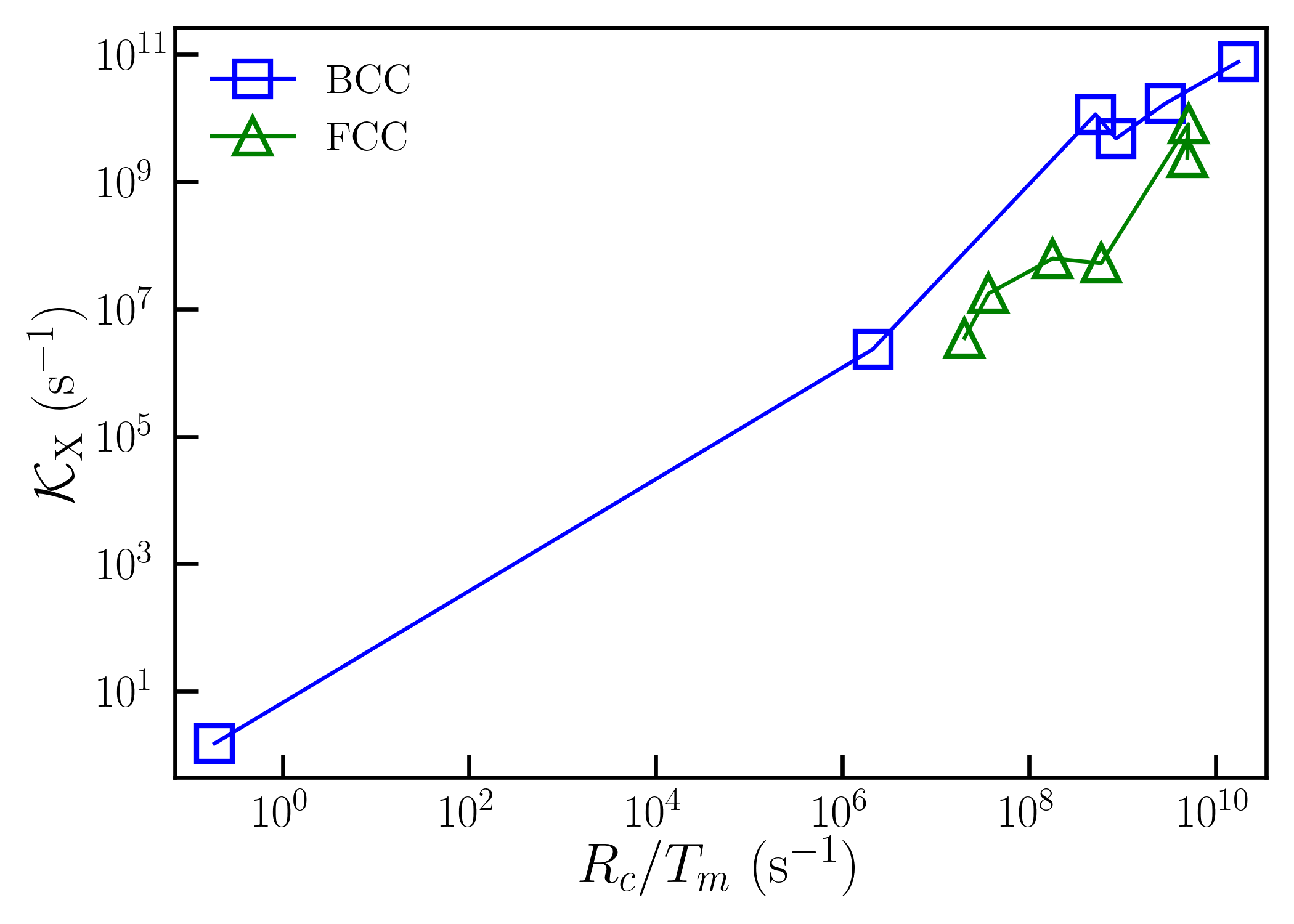}
    \caption{Characteristic crystallization rate  $\mathcal{K}_{\rm X}=(IU^3)^{1/4}$ at $T=0.8T_m$ plotted as a function of the critical cooling rate $R_c$ divided by the melting temperature $T_m$ for metallic liquids that crystallize into BCC/B2 (squares) and FCC phases (triangles).}
    \label{fig:Kx}
\end{figure}

In Fig.~\ref{fig:Kx}, we show a strong positive correlation between $\mathcal{K}_{\rm X}$ and the critical cooling rate $R_c$ for the $12$ metallic liquids that we considered, with both quantities spanning approximately eleven orders of magnitude. 
Systems with smaller $R_c$ consistently have smaller $\mathcal{K}_{\rm X}$, indicating that slower nucleation and growth processes give rise to longer crystallization time scales. We find that $\mathcal{K}_{\rm X} \sim R_c$ for metallic liquids with both equilibrium BCC/B2 and FCC crystal structures.

The correlation between $\mathcal{K}_{\rm X}$ and $R_c$ reflects the competition between the cooling and crystallization time scales. 
The critical cooling rate specifies how rapidly a liquid must be cooled to avoid crystallization, whereas $\mathcal{K}_{\rm X}$ characterizes the rate at which crystals form at a representative undercooled temperature. 
A smaller $\mathcal{K}_{\rm X}$  corresponds to a longer crystallization time and therefore allows slower cooling without crystallization. In contrast, a larger $\mathcal{K}_{\rm X}$ implies faster crystallization, which requires more rapid cooling to achieve glass formation. Continuous cooling of the metallic liquids samples a range of temperatures, and thus evaluating $\mathcal{K}_{\rm X}$ at a single temperature does not fully describe the crystallization process during continuous cooling. Nevertheless, the strong correlation in Fig.~\ref{fig:Kx} between $\mathcal{K}_{\rm X}$ and $R_c$ indicates that studying the nucleation and growth processes at $0.8T_m$ can approximate the crystallization process during cooling. The benefit of this approach is that the GFA can be estimated without carrying out direct MD simulations of cooling even for good glass formers with $R_c < 10^7$ K/s.

\subsection{Kinetic and thermodynamic contributions to the crystallization rate}
\label{sec:resultB}

Having established the connection between the characteristic crystallization rate and GFA, we next separate the kinetic and thermodynamic contributions to $\mathcal{K}_{\rm X}$ to identify the contribution that controls the large variation in $\mathcal{K}_{\rm X}$.  The critical size of a spherical nucleus within CNT is 
\begin{equation}
    N_c=\frac{32\pi{\gamma}^3}{3\rho_s^2\Delta\mu^3},
    \label{eq:Nc}
\end{equation}
where $\rho_s$ is the number density of the crystal phase and $\gamma$ is the effective liquid--crystal interfacial free energy. Equation~\ref{eq:Nc} allows us to evaluate $\gamma$ from $N_c$, $\rho_s$, and $\Delta\mu$.

\begin{figure}[t]
    \centering
    \includegraphics[width=0.95\linewidth]{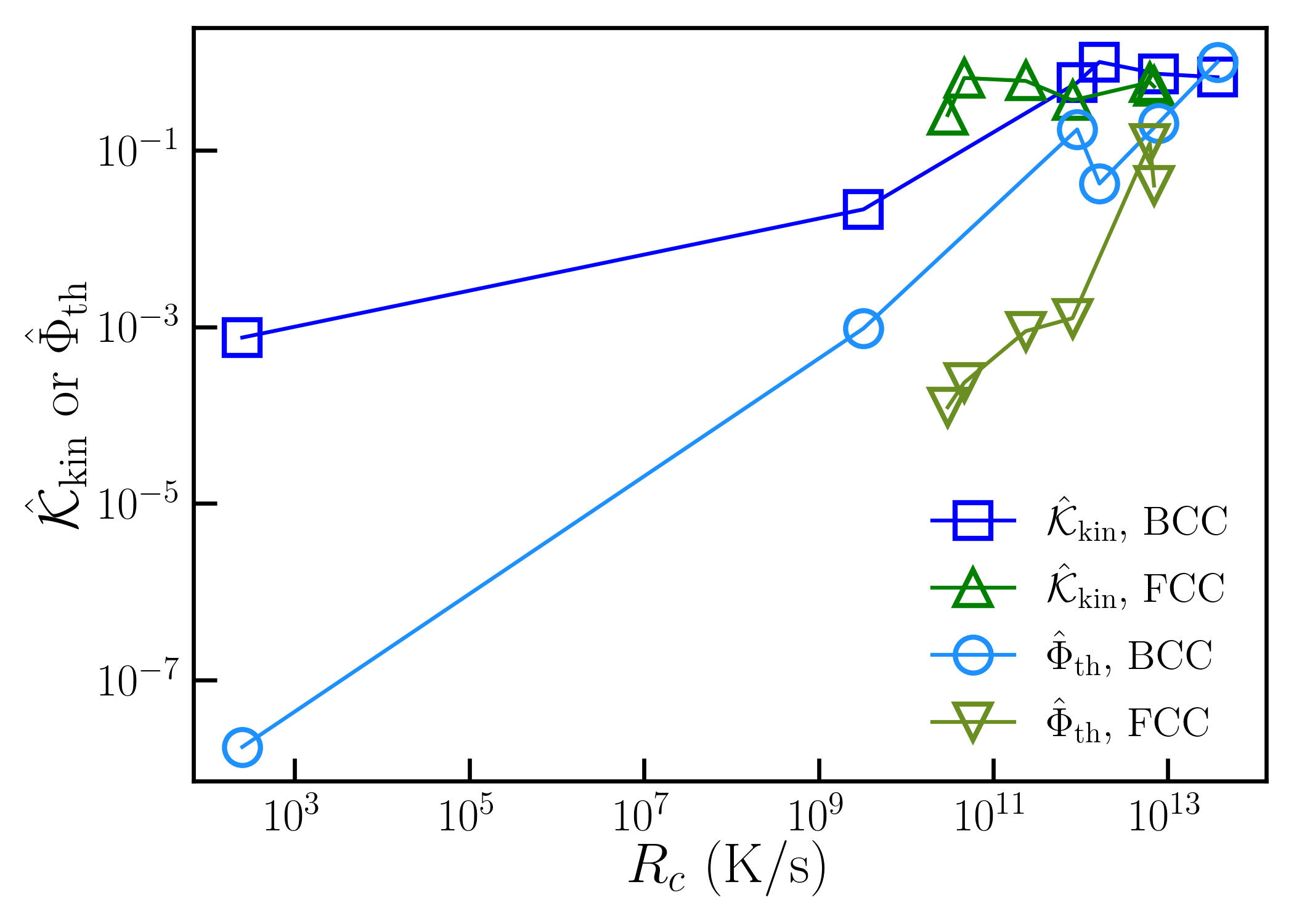}
    \caption{Normalized kinetic $\hat{\mathcal{K}}_{\rm kin}$ and thermodynamic $\hat{\Phi}_{\rm th}$ contributions to the crystallization rate $\mathcal{K}_{\rm X}$ plotted as functions of $R_c$. The kinetic contribution spans approximately three orders of magnitude, whereas the thermodynamic contribution spans approximately eight orders of magnitude.}
    \label{fig:KinThermo}
\end{figure}

After introducing the dimensionless thermodynamic quantities
\begin{subequations}
\begin{align}
    \Delta\tilde{\mu}&=\beta\Delta\mu\\
    \intertext{and} 
    \tilde{\gamma}&=\frac{\beta\gamma}{\rho_s^{2/3}},
    \label{eq:reduced}
\end{align}
\end{subequations}
and substituting Eqs.~\ref{eq:I}--\ref{eq:Nc} into
Eq.~\ref{eq:KX}, the crystallization rate can be decomposed into the product of two factors 
\begin{equation}
    \mathcal{K}_{\rm X}
    =
    \mathcal{K}_{\rm kin}\Phi_{\rm th},
    \label{eq:decomposition}
\end{equation}
where
\begin{equation}
    \mathcal{K}_{\rm kin}
    =
    \left[\rho_l f^+(k_0D)^3\right]^{1/4}
    \label{eq:Kkin}
\end{equation}
is the kinetic contribution with units of s$^{-1}$ and
\begin{equation}
    \Phi_{\rm th}
    =
    \frac{\Delta\tilde{\mu}^{1/2}}
    {(8\pi)^{1/4}\tilde{\gamma}^{3/8}}
    \left(1-e^{-\Delta\tilde{\mu}}\right)^{3/4}
    \exp\left[
    -\frac{4\pi\tilde{\gamma}^3}
    {3\Delta\tilde{\mu}^2}
    \right]
    \label{eq:Phi}
\end{equation}
is the dimensionless thermodynamic contribution. Although $\rho_l$ is not intrinsically a kinetic quantity, it is included in $\mathcal{K}_{\rm kin}$ so that $\mathcal{K}_{\rm kin}$ has units of inverse time. The ratio of the maximum to minimum values of $\rho_l$, $\max(\rho_l)/\min(\rho_l)\approx 2.15$, over the $12$ metallic liquids that we studied.
Equation~\ref{eq:decomposition} separates the kinetic contributions that are set by the diffusion coefficient and atomic attachment rate from the thermodynamic contributions to the nucleation and growth processes. $\Phi_{\rm th}$ includes both the nucleation free-energy barrier and the bulk driving force. 

To enable comparisons between $\mathcal{K}_{\rm kin}$ and $\Phi_{\rm th}$, we normalize them by the maximum values over the metallic liquids that we studied, 
\begin{equation}
    \hat{\mathcal{K}}_{\rm kin}
    =
    \frac{\mathcal{K}_{\rm kin}}
    {\max(\mathcal{K}_{\rm kin})},
    \qquad
    \hat{\Phi}_{\rm th}
    =
    \frac{\Phi_{\rm th}}
    {\max(\Phi_{\rm th})}.
    \label{eq:normalized}
\end{equation}
As shown in Fig.~\ref{fig:KinThermo}, both $\hat{\mathcal{K}}_{\rm kin}$ and $\hat{\Phi}_{\rm th}$ (and $\mathcal{K}_{\rm kin}$ and $\Phi_{\rm th}$) decrease with $R_c$, but their ranges differ markedly. For the metallic liquids that we studied, $\mathcal{K}_{\rm kin}$ spans approximately three orders of magnitude, whereas $\Phi_{\rm th}$ spans approximately eight orders of magnitude. 
Since $\ln\mathcal{K}_{\rm X}=\ln\mathcal{K}_{\rm kin}+\ln\Phi_{\rm th}$, the larger variation in $\Phi_{\rm th}$ indicates that the thermodynamic contribution more strongly influences $\mathcal{K}_{\rm X}$ than the kinetic contribution. The next step is to identify which thermodynamic factors give rise to large variations in $\Phi_{\rm th}$.

\subsection{Factors that control nucleation barrier}
\label{sec:resultC}

We now investigate the thermodynamic contribution $\Phi_{\rm th}$ to $\mathcal{K}_{\rm X}$ and why it can vary so strongly over the metallic liquids that we studied. Equation~\ref{eq:Phi} for the thermodynamic contribution can be written as a product of three factors, $\Phi_{\rm th}=\phi_{{\rm th,I}} \phi_{{\rm th,II}} \phi_{{\rm th,III}}$: 
\begin{equation}
    \Phi_{\rm th}=
    \underbrace{
        \frac{\Delta\tilde\mu^{1/2}}
        {(8\pi)^{1/4}\tilde\gamma^{3/8}}
    }_{\phi_{\rm th,I}}
    \underbrace{
        \left(1-e^{-\Delta\tilde\mu}\right)^{3/4}
        \vphantom{\frac{\Delta\tilde\mu^{1/2}}
        {(8\pi)^{1/4}\tilde\gamma^{3/8}}}
    }_{\phi_{\rm th,II}}
    \underbrace{
        \exp\left[-\frac{\beta \Delta G_c}{4}\right]
        \vphantom{\frac{\Delta\tilde\mu^{1/2}}
        {(8\pi)^{1/4}\tilde\gamma^{3/8}}}
    }_{\phi_{\rm th,III}},
    \label{eq:Phi_parts}
\end{equation}
where the reduced nucleation free-energy barrier
\begin{equation}
    \beta \Delta G_c = \frac{16\pi\tilde\gamma^3}{3\Delta\tilde\mu^2}.
    \label{eq:reducedbarrier}
\end{equation}
The first factor stems from the Zeldovich prefactor in the nucleation rate, the second describes the thermodynamic driving force in the crystal growth rate, and the third represents the exponential suppression associated with the nucleation free-energy barrier. 
Across the metallic liquids that we studied, the ratios of the maximum to minimum values of $\phi_{\rm th,I}$ and $\phi_{\rm th,II}$ are small, i.e., $\sim 1.32$ and $1.96$, respectively.
Their small variation cannot explain the approximately eight-order-of-magnitude range of $\Phi_{\rm th}$. Instead, the dominant variation arises from $\phi_{\rm th,III}$, i.e.
\begin{equation}
    \ln \phi_{\rm th,III} =-\frac{4\pi}{3}
    \frac{\tilde\gamma^3}{\Delta\tilde\mu^2}.
    \label{eq:phi3var}
\end{equation}
Thus, the dimensionless nucleation barrier depends on the product $\tilde\gamma^3\Delta\tilde\mu^{-2}$. A larger reduced interfacial free energy increases the barrier, whereas a stronger reduced driving force lowers it. 
At $T=0.8T_m$, the approximation $\Delta\mu(T)=\Delta H_m(1-T/T_m)$ gives $\Delta\tilde\mu=\Delta H_m/(4k_{\rm B}T_m)$. Thus, we assume that the reduced bulk driving force is controlled by the reduced enthalpy of fusion.

\begin{figure}[t]
    \centering
    \includegraphics[width=0.95\linewidth]{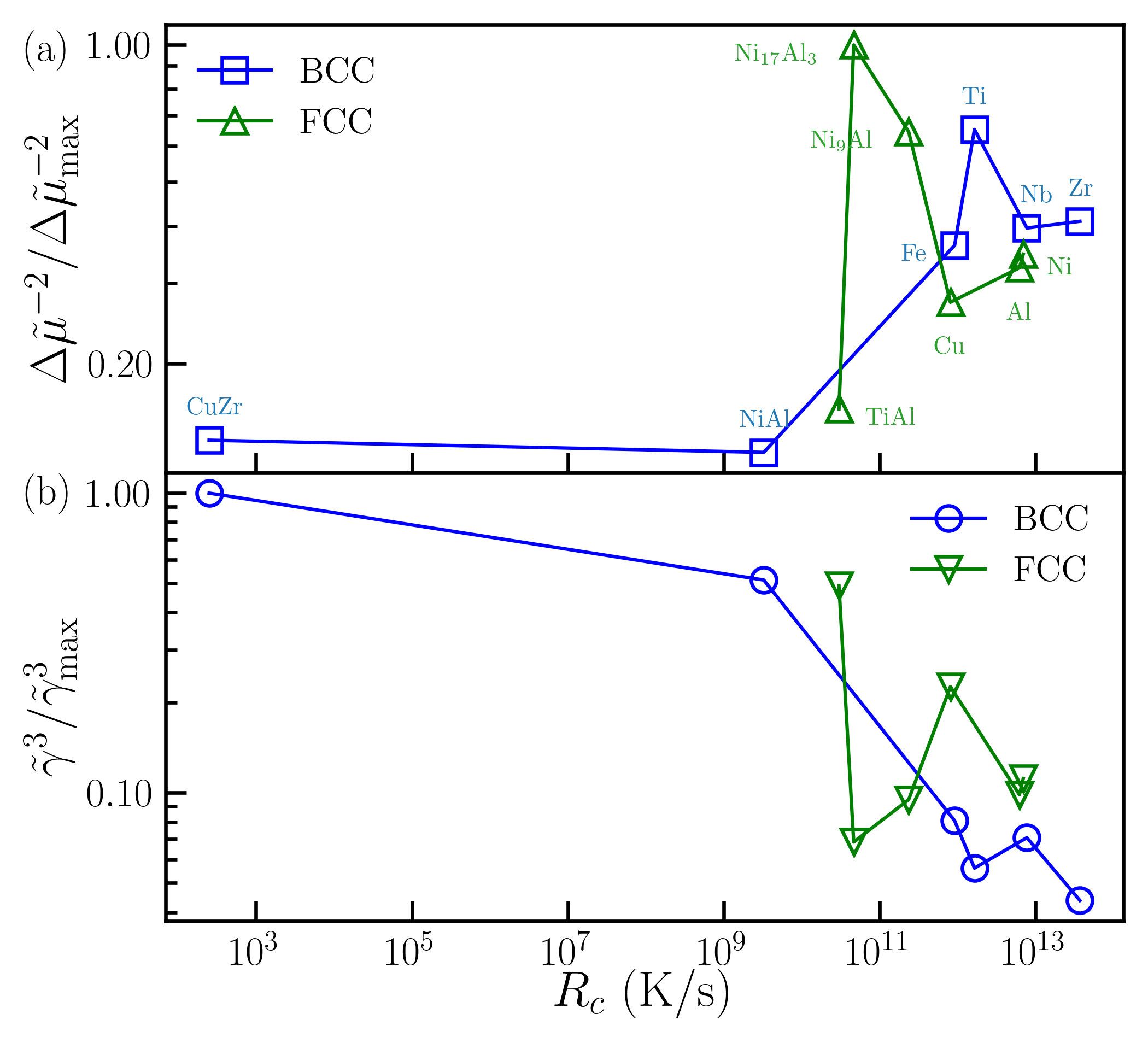}
    \caption{Two factors contribute to the dimensionless nucleation barrier $\beta \Delta G_c$ in Eq.~\ref{eq:reducedbarrier}. (a) Inverse squared reduced driving force, $\Delta\tilde\mu^{-2}$, normalized by its maximum and (b) the cubed reduced effective liquid--crystal interfacial free energy, $\tilde\gamma^3$, normalized by its maximum, plotted versus $R_c$.}
    \label{fig:DmuGam}
\end{figure}

In Fig.~\ref{fig:DmuGam}, we plot $\Delta\tilde\mu^{-2}$ and $\tilde\gamma^3$, each normalized independently by its maximum value, as a function of $R_c$. For CuZr, NiAl, and TiAl, the comparatively larger interfacial free-energy term gives rise to a high nucleation barrier despite a relatively strong bulk driving force for crystallization that tends to lower it.
In contrast, Ni$_9$Al, Ni$_{17}$Al$_3$, and Ti possess a weak driving force, which causes large values of $\Delta\tilde\mu^{-2}$ and a relatively large nucleation barrier. In pure Cu, the barrier-enhancing effect of a relatively large interfacial free energy is partly offset by a strong driving force. The remaining pure metals have comparable values for the driving force, so differences in the effective interfacial free energy more directly explain variations in the nucleation barrier.

These results establish two complementary thermodynamic mechanisms for suppressing crystallization, i.e., increasing the effective liquid--crystal interfacial free energy and reducing the bulk driving force. In general, the relative sizes of the interfacial free energy and bulk driving force differ across metallic liquids, and these factors can coordinate to enhance the nucleation barrier ($\Delta\tilde\mu^{-2}$ and $\tilde\gamma^3$ increase), coordinate to decrease the nucleation barrier ($\Delta\tilde\mu^{-2}$ and $\tilde\gamma^3$ decrease), or compete so that they offset one another.

\subsection{Local structural and chemical signatures of crystallization resistance}
\label{sec:resultD}

To explore the microscopic origins for increases in the crystal nucleation barrier, we calculate structural frustration within the supercooled liquid and the chemical-order mismatch between the liquid and crystal. To quantify structural frustration, we calculate the frequency of icosahedral- and crystal-like local environments within the liquid. For the chemical order mismatch, we compare the fraction of neighboring atoms of each type for the liquid and crystal phases.

We define the structural frustration as
\begin{equation}
    S_{\rm frust} = (1-F_{\rm cry})F_{\rm ico},
    \label{eq:Sfrust}
\end{equation}
where $F_{\rm cry}$ and $F_{\rm ico}$ are the fractions of atoms participating in crystal- and icosahedral-like local environments, respectively, in the supercooled liquid at $T=0.8T_m$. We define these environments based on bond-orientational order and Voronoi indices (see Appendix~\ref{app:order}). 
Large $S_{\rm frust}$ values indicate abundant icosahedral ordering with relatively little crystal-like ordering. Conversely, a small value can reflect either few icosahedral local environments or appreciable crystalline preordering, or both. Icosahedral order is incompatible with the translational periodicity of the crystal structures considered here, and therefore it frustrates the formation and growth of crystalline order.

\begin{figure}[t]
    \centering
    \includegraphics[width=0.95\linewidth]{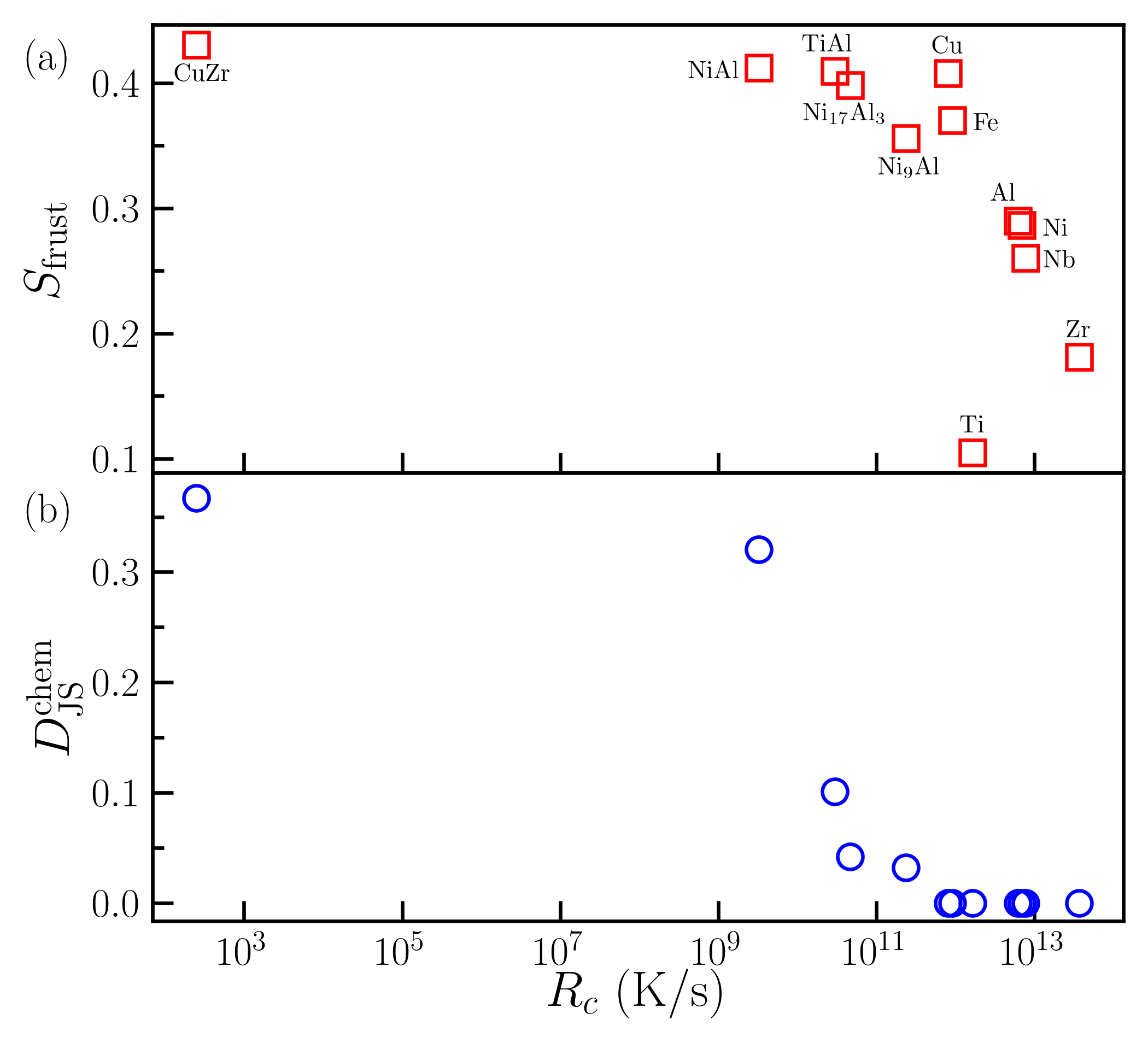}
    \caption{(a) Structural frustration $S_{\rm frust}=(1-F_{\rm cry})F_{\rm ico}$, defined using the fractions of atoms with crystal- and icosahedral-like local environments within the supercooled liquid at $0.8T_m$. (b) Chemical-order mismatch $D_{\rm JS}^{\rm chem}$, calculated using the Jensen--Shannon divergence between local chemical-order distributions in the liquid and crystal. Good glass formers exhibit stronger structural frustration and larger chemical-order mismatch.}
    \label{fig:frust}
\end{figure}

As shown in Fig.~\ref{fig:frust} (a), better glass formers generally exhibit larger $S_{\rm frust}$. CuZr, the best glass former among the metallic liquids that we studied, has one of the largest $S_{\rm frust}$ values. 
Notably, CuZr and NiAl have comparable fractions of crystal-like atoms, whereas the remaining systems, which are poorer glass formers, have $F_{\rm cry}$ values that are $\sim 2.53$--$4.89$ times larger. This pronounced difference highlights the importance of crystalline preordering in the supercooled liquid, since it can facilitate subsequent crystallization. 
In addition, the increased fraction of locally favored noncrystalline structures in better glass formers can frustrate the development of crystalline order. Nevertheless, because $S_{\rm frust}$ depends on both $F_{\rm cry}$ and $F_{\rm ico}$, variations in $S_{\rm frust}$ reflect the combined effects of crystalline preordering and icosahedral ordering. 

To characterize the chemical-order mismatch, we compare distributions of the local chemical ordering for each atom type in the supercooled liquid and crystal. We define the chemical-order mismatch as 
\begin{equation}
    D_{\rm JS}^{\rm chem} =  \frac{1}{n}\sum_{\alpha=1}^{n} D_{\rm JS}(P_l^{(\alpha)}, P_s^{(\alpha)}),
    \label{eq:DJchem}
\end{equation}
where $n$ is the number of atom types in the system, and $P_l^{(\alpha)}$ and $P_s^{(\alpha)}$ are the local chemical-order distributions of $\alpha$-type atoms in the liquid and crystal, respectively. Calculations of the chemical order are provided in Appendix~\ref{app:order}. The Jensen--Shannon divergence $D_{\rm JS}(P_l, P_s)$ between distributions $P_l$ and $P_s$ is ~\cite{manning1999, zhang2023local}
\begin{equation}
    D_{\rm JS}(P_l, P_s)  =
    \frac{1}{2}D_{\rm KL}(P_l, M)
    +
    \frac{1}{2}D_{\rm KL}(P_s, M),
    \label{eq:DJS}
\end{equation}
where $M=(P_l+P_s)/2$, and $D_{\rm KL}$ denotes the Kullback--Leibler divergence,
\begin{equation}
    D_{\rm KL}(P, M) = \int P(x)\log_2\! \left[\frac{P(x)}{M(x)}\right] dx.
    \label{eq:KL}
\end{equation}
For distributions defined on discrete bins, the integral is replaced by a sum. With base-2 logarithms, $0\leq D_{\rm JS}^{\rm chem}\leq1$, where zero denotes identical distributions and larger $D_{\rm JS}^{\rm chem}$ indicates greater chemical-order mismatches. For pure metals, the local chemical ordering of each atom is, by definition, always one, so $D_{\rm JS}^{\rm chem}\equiv 0$.

In Figure~\ref{fig:frust} (b), we show that the chemical-order mismatch generally increases with decreasing $R_c$. In particular, CuZr and NiAl exhibit relatively large $D_{\rm JS}^{\rm chem}$, whereas Ni$_9$Al, Ni$_{17}$Al$_3$, TiAl, and the pure metals have smaller values. 
The larger mismatch in these better glass formers suggests that atomic attachment to a growing crystal may require more extensive rearrangements of both atomic positions and local chemical environments. Conversely, greater similarity between the chemical environments of the liquid and crystal may accelerate atomic cluster attachment and promote crystal growth. Consistent with this interpretation, systems with small chemical-order mismatches generally exhibit relatively large $\mathcal{K}_{\rm kin}$. These observations suggest a possible microscopic connection between chemical-order mismatches and the kinetic contribution to the crystallization rate shown in Fig.~\ref{fig:KinThermo}. 

The structural and chemical order metrics also provide insights into variations in the effective interfacial free energy.
Forming a crystalline nucleus in a liquid rich in noncrystalline local environments may require substantial structural reorganization, while a chemical-order mismatch between the liquid and crystal can further increase the interfacial free energy. In CuZr, strong structural frustration and substantial chemical-order mismatch accompany a relatively large effective interfacial free energy, which suggests that both structural frustration and chemical-order mismatches contribute to the nucleation barrier.

The Ni$_9$Al and Ni$_{17}$Al$_3$ metallic liquids illustrate another connection between local atomic ordering and thermodynamics. Their relatively small chemical-order mismatch is consistent with their chemically disordered FCC crystal structures. Unlike CuZr, however, their nucleation barriers are enhanced primarily by weak bulk thermodynamic driving forces rather than large effective interfacial free energies. Similar local chemical environments in the liquid and crystal may lead to comparable bonding energies, while crystalline preordering in the liquid may reduce the structural rearrangements and associated energetic changes during crystallization. Both effects may contribute to a smaller enthalpy of fusion, $\Delta H_m$, and consequently a weaker thermodynamic driving force. Together, these observations suggest that local structural and chemical ordering can influence crystallization resistance through separate thermodynamic mechanisms, either by increasing the effective interfacial free energy or by reducing the bulk driving force.

\section{Conclusions and future directions}
\label{sec:conclusions}

We have quantified how crystal nucleation and growth, atomic kinetic processes, and thermodynamic mechanisms contribute to GFA in $12$ model metallic systems with BCC/B2 and FCC-based crystal structures. We emphasize three key findings. First, the characteristic crystallization rate $\mathcal{K}_{\rm X}=(IU^3)^{1/4}$, evaluated at $T=0.8T_m$, is strongly correlated with the critical cooling rate $R_c$ over approximately eleven orders of magnitude. This relationship links the combined nucleation and growth kinetics at a representative undercooling to the large variation in GFA across a broad range of metallic systems.

Second, we decompose $\mathcal{K}_{\rm X}$ into kinetic and thermodynamic contributions, where the kinetic factor gives rise to a variation of three orders of magnitude, whereas the thermodynamic factor gives rise to eight. Therefore, the thermodynamic contribution accounts for the larger share of the variation in $\mathcal{K}_{\rm X}$. Further, we show that the dominant thermodynamic variation is caused by the crystal nucleation barrier.

Within CNT, the nucleation barrier depends on the product $\tilde\gamma^3 \Delta\tilde\mu^{-2}$. In CuZr, NiAl, and TiAl, a relatively large effective interfacial free energy favors a high nucleation barrier, whereas weak bulk driving forces occur for Ni$_9$Al, Ni$_{17}$Al$_3$, and Ti. The interfacial free energy and driving force can also partially compensate. Thus, both the interfacial free energy and driving force are needed to determine the thermodynamic contribution to crystallization resistance for metallic liquids.

Structural frustration within the supercooled liquid and chemical-order mismatches between the liquid and crystal can provide microscopic insights into crystallization resistance. The greater structural frustration and chemical-order mismatch generally observed in better glass formers suggest that crystallization requires more extensive atomic rearrangements, potentially increasing the effective interfacial free energy. Conversely, similarities in local structural and chemical environments between the liquid and crystal may contribute to weaker bulk driving forces in some alloys. Together, these findings establish a quantitative framework for assessing the relative importance of the kinetic and thermodynamic contributions to the GFA and identifying distinct thermodynamic mechanisms that suppress crystallization.

We identify several promising future directions for this research. First, determining $I(T)$ and $U(T)$ over a broader temperature range could assess the predictive accuracy of the crystallization rate and enable direct integration of nucleation and growth kinetics during continuous cooling. Second, incorporating crystal nucleus shape and interfacial anisotropy could establish a more detailed relationship between the effective $\gamma$ inferred from CNT and orientation-dependent liquid--crystal interfacial free energies. Third, examining heterogeneous nucleation and the influence of pre-existing crystalline seeds could clarify their effects on crystallization kinetics and improve GFA prediction. Finally, extending the analysis to a wider range of alloy compositions and crystalline phases would test the generality of the kinetic and thermodynamic decomposition of the crystallization rate and its potential for predicting glass-forming alloys.

\begin{acknowledgments}
We acknowledge support from NSF Grant No. CMMI-1901959 (J.F., W.J., and C.S.O.). This work was also supported by the High Performance Computing facilities operated by Yale's Center for Research Computing.
\end{acknowledgments}

\section*{DATA AVAILABILITY}
The data that support the findings of this article are openly available~\cite{data}; embargo periods may apply.

\appendix
\section{Determination of the melting temperature and enthalpy of fusion}
\label{app:Tm}

We determine the melting temperature $T_m$ of each pure metal or alloy using a solid--liquid coexistence method in which a portion of the initial crystal is temporarily pinned~\cite{kerrache2008molecular,tang2013anomalously}. The initial simulation cell contains $10\times10\times30$ conventional unit cells for FCC/L1$_0$ structures and $13\times13\times39$ conventional unit cells for BCC/B2 structures, corresponding to 12,000 and 13,182 atoms, respectively. Periodic boundary conditions are imposed in all three spatial directions.

Initially, atoms in the central third of the simulation cell along its long axis are pinned, while the remaining atoms are heated from room temperature to 1500 K above the experimental melting temperature at a rate of $5\times10^{11}$ K/s. The resulting liquid region is equilibrated for 200 ps and then cooled to a prescribed temperature at the same rate. During this preparation process, the simulation cell is maintained at zero external pressure, with the transverse dimensions $L_x$ and $L_y$ coupled and the box length along the long axis allowed to vary independently. The positions of the pinned atoms are scaled so that they mimic changes in the simulation cell.
After the preparation is completed, the pinned atoms are released, and the system is relaxed at the prescribed temperature for 10 ps, followed by a 1 ns $NpT$ simulation during which the atomic volume is monitored. 

\begin{figure}[t]
    \centering
    \includegraphics[width=0.95\linewidth]{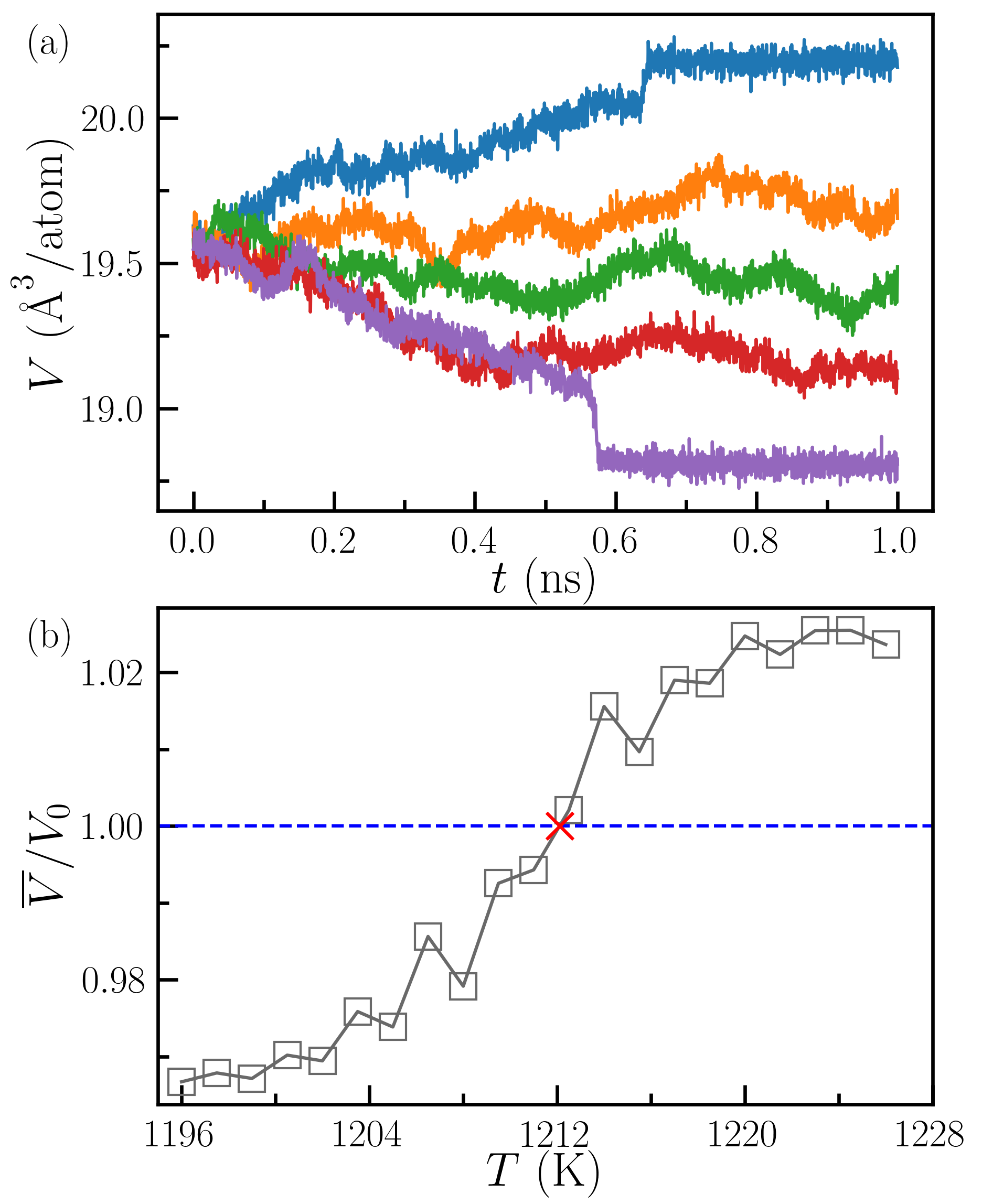}
    \caption{Determination of the melting temperature of Al using the solid--liquid coexistence simulations. (a) Atomic volume plotted as a function of time at prescribed temperatures $T=1217$, $1212.5$, $1211$, $1206.5$, and $1203.5$ K, ordered from the uppermost to the lowermost curve at $t=1$ ns. (b) Ratio of the time-averaged atomic volume $\overline{V}$ to its initial value $V_0$ plotted as a function of temperature. $\overline{V}/V_0=1$ determines the melting temperature $T_m\approx1212$ K.}
    \label{fig:Tm}
\end{figure}

Below $T_m$, crystal growth reduces the system volume, whereas above $T_m$, melting increases it, as illustrated in Fig.~\ref{fig:Tm} (a). We therefore identify $T_m$ from the temperature at which the volume-change ratio $\overline{V}/V_0=1$, where $V_0$ is the atomic volume at the start of the final 1-ns simulation and $\overline{V}$ is the time-averaged atomic volume over that interval. In Figure~\ref{fig:Tm} (b), we show this ratio for Al, yielding $T_m\approx1212$ K.

The thermodynamic driving force, i.e., the chemical-potential difference between the liquid and crystal, is estimated using $\Delta\mu(T)=\Delta H_m\left(1-{T}/{T_m}\right)$,
where $\Delta H_m$ is the enthalpy of fusion per atom at $T_m$. We have verified that this approximation gives results close to those obtained by thermodynamic integration from coexistence.
To calculate $\Delta H_m$, we prepare separate crystal and liquid configurations at $T_m$. The crystal contains $14\times14\times14$ conventional unit cells for FCC/L1$_0$ structures and $18\times18\times18$ conventional unit cells for BCC/B2 structures; for Ni$_9$Al and Ni$_{17}$Al$_3$, we instead use $14\times14\times15$ cells to match their compositions. 
The crystal is heated from room temperature to $T_m$ at $10^{11}$ K/s and equilibrated for 1 ns to calculate its enthalpy. It is then heated to $T_m+1500$ K at the same rate and held for 4 ns to form a liquid. The liquid is cooled to $T_m$ at $10^{11}$ K/s and equilibrated for another 1 ns to determine its enthalpy. All stages are performed in the $NpT$ ensemble at zero pressure. The enthalpy of fusion is obtained as the difference between the liquid and crystal enthalpies per atom, $\Delta H_m=h_l(T_m)-h_s(T_m)$, where $h_\lambda=E+pV$ includes the per-atom internal energy $E$ (both kinetic and potential energy) and pressure-volume work $pV$, for the liquid ($\lambda=l$) and crystal ($\lambda=s$), respectively.

\section{Structural and chemical order}
\label{app:order}

We characterize the local structural order using the bond-orientational order (BOO) parameter~\cite{steinhardt1983} and Voronoi indices~\cite{nawano2023characterizing}. 
The $l$-fold local BOO parameter of atom $i$ is defined as
\begin{equation}
q_{lm}(i) = \sum_{j=1}^{N_i}\frac{A_{ij}}{A_{tot}^i}Y_{lm}(\theta( \textbf r_{ij}),\phi( \textbf r_{ij})),
\end{equation}
where $N_i$ is the number of Voronoi neighbors of atom $i$~\cite{voronoi}, $A_{ij}$ is the area of the Voronoi face shared by atom $i$ and its Voronoi neighbor $j$, and $A_{\rm tot}^i=\sum_j A_{ij}$ is the total surface area of the Voronoi polyhedron of atom $i$. $Y_{lm}(\theta( {\vec r}_{ij}),\phi( {\vec r}_{ij}))$ is the spherical harmonic function of degree $l$ and order $m$, $\theta({\vec r}_{ij})$ and $\phi({\vec r}_{ij})$ are the polar and azimuthal angles that parameterize the orientation of the vector ${\vec r}_{ij}$ connecting atoms $i$ and $j$. 

We average the local $q_{lm}(i)$ over atom $i$ and its Voronoi neighbors,
\begin{equation}
Q_{lm}(i) = \frac{1}{N_i+1}\left(q_{lm}(i)+\sum_{j=1}^{N_i}q_{lm}(j)\right).
\end{equation}
The corresponding coarse-grained $l$-fold BOO parameter $Q_l$ is:
\begin{equation}
Q_l(i) = \sqrt{\frac{4\pi}{2l+1}\sum_{m=-l}^l|Q_{lm}(i)|^2}.
\end{equation}
We also calculate the third-order rotational invariant $w_6$, 
\begin{equation}
    w_6=\sum_{m_1+m_2+m_3=0}
    \begin{pmatrix}
    6 & 6 & 6 \\
    m_1 & m_2 & m_3
    \end{pmatrix}
    q_{6m_1}q_{6m_2}q_{6m_3},
\end{equation}
where $\begin{pmatrix}
    6 & 6 & 6 \\
    m_1 & m_2 & m_3
    \end{pmatrix}$
denotes the Wigner 3-j symbol. 

We also describe the local structure using the Voronoi indices $(n_3,n_4,n_5,n_6,\ldots)$, where $n_k$ counts the number of faces with $k$ edges in the radical Voronoi polyhedron. 
An atom is identified as the center of an icosahedral-like structure if it satisfies either the combined criteria $Q_6<0.25$ and $w_6<-0.023$ or the Voronoi criteria $n_5\geq8$ and $\sum_{k=3}^{\infty}n_k=12$. The fraction $F_{\rm ico}$ includes atoms identified as the centers of these icosahedral structures and their corresponding neighbors.

To characterize crystalline atoms,  we calculate the normalized correlation of $q_{6m}$ as 
\begin{equation}
    s_6(i,j)=\frac{\sum_{m=-6}^{6}q_{6m}(i)q_{6m}^*(j)}{\sqrt{\sum_{m=-6}^{6}|q_{6m}(i)|^2}\sqrt{\sum_{m=-6}^{6}|q_{6m}(j)|^2}},
\end{equation}
where the asterisk denotes complex conjugation and $j$ is the Voronoi neighbor of atom $i$.  A bond is classified as crystalline when $s_6(i,j)>0.7$. An atom with more than ten crystalline bonds is identified as the center of a crystal-like local structure, i.e., a crystalline atom. The fraction $F_{\rm cry}$ of crystal-like atoms includes these centers and their Voronoi neighbors, while the crystalline fraction $F_{\rm xtal}$, used in Eq.~\ref{eq:Rc_fit} for determining $R_c$, only counts the centers~\cite{hu2020physical}.

For an atom of species $\alpha$, we define its local chemical-order parameter $c_{\alpha}$ as the fraction of Voronoi neighbors belonging to the same species $\alpha$. The resulting distributions of $c_{\alpha}$ in the liquid and crystal are used to quantify the chemical-order mismatch using Eq.~\ref{eq:DJchem}.

\begin{figure}[t]
    \centering
    \includegraphics[width=0.95\linewidth]{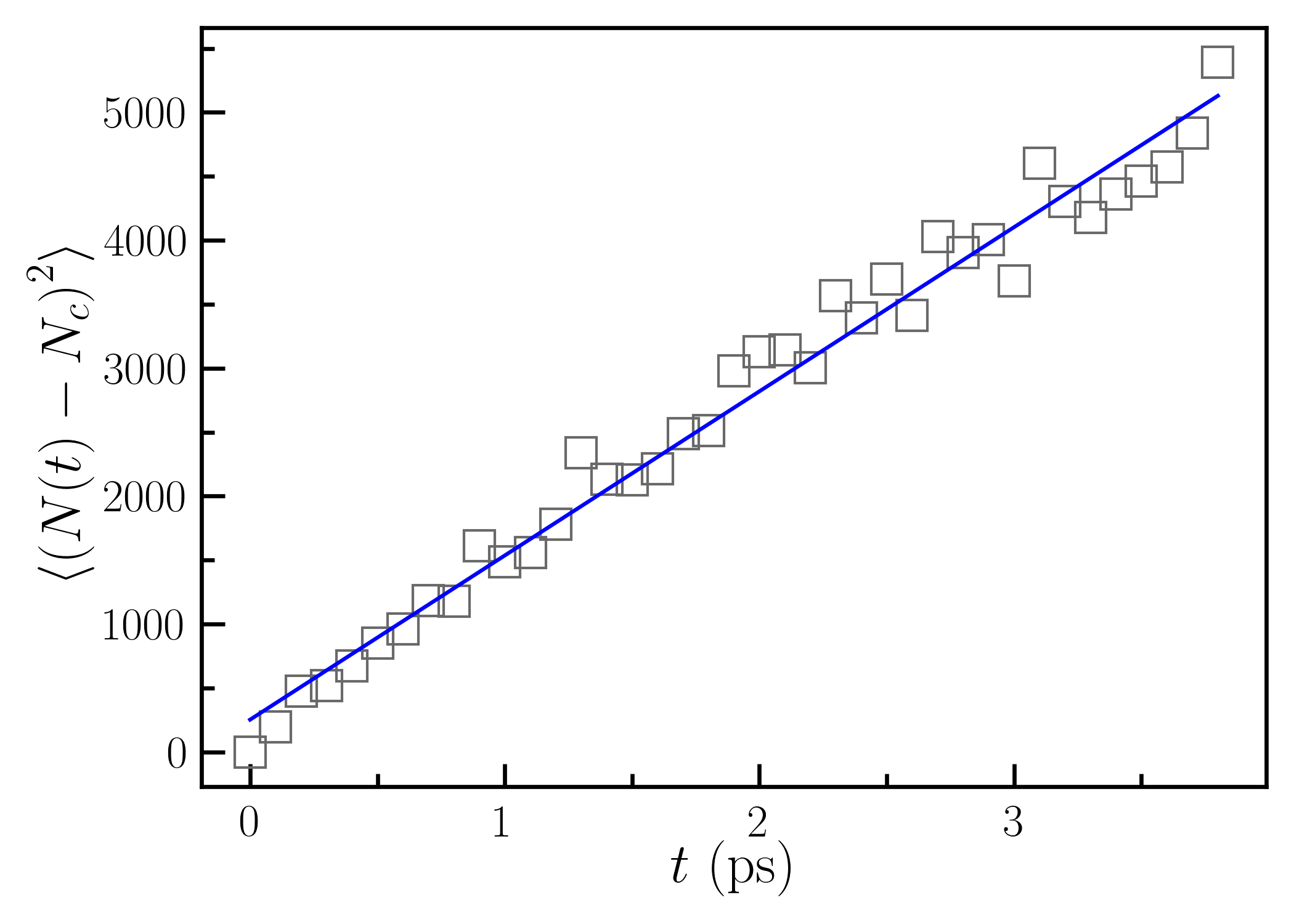}
    \caption{Mean-square fluctuations of the crystal nucleus size, $\langle[N(t)-N_c]^2\rangle$, plotted as a function of time for Fe at $T=0.8T_m$, with critical nucleus size $N_c=137$. The data are averaged over 50 independent trajectories. The linear fit has a slope of $1283.7$ ps$^{-1}$, giving an attachment rate of $f^+=641.85$ ps$^{-1}$.}
    \label{fig:fplus}
\end{figure}

\section{Seeding method for calculating cluster attachment rate}
\label{app:seed}

We determine the critical nucleus size $N_c$ at $T=0.8T_m$ using the seeding method~\cite{espinosa2015crystal}. A crystalline seed of prescribed size is embedded in a supercooled liquid, and its subsequent growth or dissolution is monitored. Seeds smaller than the critical size tend to dissolve, whereas larger seeds tend to grow. At the critical size, the probabilities of growing and dissolving are approximately equal. 

For each metallic liquid, we first construct a perfect crystal with the appropriate composition and equilibrium lattice constant at $0.8T_m$ in a simulation cell of the same size as that used for the enthalpy-of-fusion calculations. 
Atoms within a sphere of prescribed radius are pinned to form the crystalline seed, while the surrounding atoms are heated to form a liquid. The seed is then released, and the system is rapidly brought to $0.8T_m$ and relaxed for 10 ps at zero external pressure. A further 1 ns simulation at the same temperature is used to monitor the nucleus size $N(t)$, determined from the BOO-based crystalline-atom criterion described in Appendix~\ref{app:order}. We perform 50 independent trajectories for each seed size to identify the one with approximately equal probabilities of growth and dissolution.

The atomic attachment rate $f^+$ is obtained from fluctuations in the nucleus size, following Auer and Frenkel~\cite{auer2001prediction,auer2004numerical}, 
\begin{equation}
    f^+=\frac{\langle \left[N(t)-N_c\right]^2 \rangle}{2t},
\end{equation}
where $N(t)$ is the instantaneous size of the nucleus that grows or dissolves from the seed. In practice, $f^+$ is estimated from half the slope of the linear regime of $\langle[N(t)-N_c]^2\rangle$ versus $t$. Figure~\ref{fig:fplus} illustrates this procedure for Fe at $T=0.8T_m$, with $N_c=137$. A linear fit to the average over 50 trajectories yields a slope of $1283.7$ ps$^{-1}$, corresponding to $f^+=641.85$ ps$^{-1}$.

\section{Crystal growth simulations}
\label{app:growth}

We determine the crystal growth rate $U$ at $T=0.8T_m$ from MD simulations of planar solid--liquid interfaces~\cite{hu2022revealing}. The initial crystal contains $12\times12\times36$ conventional unit cells for BCC/B2 systems and $10\times10\times30$ conventional unit cells for FCC/L1$_0$ systems, constructed using the equilibrium lattice constant at $T_m$. To prepare a system at coexistence, half of the crystal is pinned while the remaining atoms are melted at $T=1.5T_m$. Periodic boundary conditions produce two planar liquid--crystal interfaces normal to the elongated $z$-direction.

The coexistence configuration is equilibrated at $T_m$ in the $Np_zT$ ensemble, with pressure $p_z=0$ held fixed by allowing the box length along its long axis to fluctuate. Crystal growth is then initiated by bringing the system to $T=0.8T_m$ in the same ensemble. To track the interfaces, the simulation cell is divided into thin slabs along the elongated direction, and the crystalline fraction $F_{\rm xtal}$ is calculated within each slab. The position of the liquid--solid interface is defined by $F_{\rm xtal}=0.5$. The crystal growth rate $U$ is obtained from the slope of the interface position as a function of time.

\end{document}